\documentclass[prl,aps,twocolumn,showpacs,superscriptaddress,footnoteinbib]{revtex4-1}

\usepackage{graphicx}% Include figure files
\usepackage{dcolumn}% Align table columns on decimal point
\usepackage{bm}% bold math
\usepackage{dsfont}
\usepackage{amssymb,amsmath}
\usepackage{sidecap}
\usepackage{wrapfig}

\usepackage{graphicx}
\usepackage{leftidx}
\usepackage{color}

\usepackage{bbold} %This package allows to write the identity symbol

\usepackage{wasysym} % In order to introduce polygon symbols in the text
\newcommand{\be}{\begin{equation}}
\newcommand{\ee}{\end{equation}}
\newcommand{\ba}{\begin{align}}
\newcommand{\ea}{\end{align}}
\newcommand{\sysb}{\left\{\begin{array}}
\newcommand{\syse}{\end{array}\right.}
\newcommand{\baa}{\begin{array}}
\newcommand{\eaa}{\end{array}}

\newcommand{\mal}{\mathcal}

\newcommand{\lt}{\left(}
\newcommand{\rt}{\right)}
\newcommand{\lqq}{\left[}
\newcommand{\rqq}{\right]}

\newcommand{\wh}{\widehat}

\newcommand{\id}{\mathbb{1}}

\newcommand{\trace}[1]{{\rm tr}\left\{ #1 \right\}}

\newcommand{\matb}{\left(\begin{array}}
\newcommand{\mate}{\end{array}\right)}

\newcommand{\abs}[1]{\left| #1 \right|}

\newcommand{\sx}{\wh{\sigma}^x}
\newcommand{\sy}{\wh{\sigma}^y}
\newcommand{\sz}{\wh{\sigma}^z}

\newcommand{\comm}[2]{\left[ #1, #2 \right]}
\newcommand{\acomm}[2]{\left\{ #1, #2 \right\}}

\newcommand{\av}[1]{\left\langle  #1  \right\rangle}

\newcommand{\reff}[1]{(\ref{#1})}

\newcommand{\suml}[2]{\sum\limits_{#1}^{#2}}

\newcommand{\Z}{\mathbb{Z}}

\newcommand{\hrho}{\wh{\rho}}

\newcommand{\Decay}{\Gamma}
\newcommand{\Deph}{K}

\begin{document}

\title{Universal non--equilibrium properties of dissipative Rydberg gases}

\author{Matteo Marcuzzi}
\affiliation{School of Physics and Astronomy, University of Nottingham, Nottingham, NG7 2RD, United Kingdom}
\author{Emanuele Levi}
\affiliation{School of Physics and Astronomy, University of Nottingham, Nottingham, NG7 2RD, United Kingdom}
\author{Sebastian Diehl}
\affiliation{Institute for Theoretical Physics, University of Innsbruck, A-6020 Innsbruck, Austria}
\author{Juan P. Garrahan}
\affiliation{School of Physics and Astronomy, University of Nottingham, Nottingham, NG7 2RD, United Kingdom}
\author{Igor Lesanovsky}%\affiliation{School of Physics and Astronomy, University of Nottingham, Nottingham, NG7 2RD, United Kingdom}
\affiliation{School of Physics and Astronomy, University of Nottingham, Nottingham, NG7 2RD, United Kingdom}

\begin{abstract}
We investigate the out-of-equilibrium behavior of a dissipative gas of Rydberg atoms that features a dynamical transition between two stationary states characterized by different excitation densities. We determine the structure and properties of the phase diagram and identify the universality class of the transition, both for the statics and the dynamics. We show that the proper dynamical order parameter is in fact not the excitation density and find evidence that the dynamical transition is in the ``model A'' universality class, i.e. it features a non-trivial $\Z_2$ symmetry and a dynamics with non-conserved order parameter. This sheds light on some relevant and observable aspects of dynamical transitions in Rydberg gases.
In particular it permits a quantitative understanding of a recent experiment [C. Carr \emph{et al.}, Phys. Rev. Lett. \textbf{111}, 113901 (2013)] which observed bistable behaviour as well as power-law scaling of the relaxation time. The latter emerges not due to critical slowing down in the vicinity of a second order transition, but from the non-equilibrium dynamics near a so-called spinodal line.
\end{abstract}

\maketitle
\textit{Introduction.---}
The study of the emergence of collective behavior in many-body systems continues to be a very active field of research. Fundamental insights, such as the onset of universality and its consequences \cite{Zinn-Justin,QPT,Ma} are central for our understanding of matter in general. In recent years, there has been a growing interest in understanding dynamical phase transitions \cite{HH,kamenevbook,tauberbook} in the context of driven open many-body quantum systems \cite{Baumann09,DPT1, DPT2, nagy11, oztop12, DPT_Diehl, esslingerdicke3, Kessler12,Foss-Feig13,DPT4,DPT3, sieberer13, tauber14}, and progress in the manipulation of ultracold atoms \cite{Bloch} has made it possible to access and explore many-body phenomena under precisely controllable experimental conditions \cite{Exp-Bose-Hubbard-3D,Greiner,Kinoshita,Light-cone-exp,Exp-Bose-Hubbard-3D,Exp-Luttinger}.

%The study of the emergence of collective behavior in many-body systems continues to be a very active field of research. Fundamental insights, such as the onset of universality and its consequences \cite{Zinn-Justin,QPT,Ma} are central for our understanding of matter in general. In recent years, there has been a growing interest in understanding dynamical phase transitions in many-body quantum systems \cite{Baumann09,DPT1, DPT2, DPT_Diehl, Kessler12,Foss-Feig13,DPT4,DPT3}, and progress in the manipulation of ultracold atoms \cite{Bloch} has made it possible to access and explore many-body phenomena under precisely controllable experimental conditions \cite{Exp-Bose-Hubbard-3D,Greiner,Kinoshita,Light-cone-exp,Exp-Bose-Hubbard-3D,Exp-Luttinger}.

In this context, a class of systems that offers a rich and intricate physics is represented by so-called Rydberg gases \cite{Schwarzkopf2011,Rydberg2, Ryd-QI, Experiment2, Experiment1,Hofmann2013}, i.e., atomic clouds in which atoms are laser-excited to high-lying energy levels. The main consequence of the population of such orbitals is a considerable increase \cite{Rydberg2, Ryd-QI} in the interaction strength. This is at the heart of several non-trivial dynamical phenomena, both for closed systems undergoing coherent evolution and showing enhanced spatial (anti-)correlations \cite{Weimer08,Pohl2010,Ryd-lattice1, Ryd-lattice2}, and for open ones, in which the interplay between driving and dissipation leads instead to intermittency \cite{Ryd-int}, glassy behavior \cite{PRL-KinC} and bistable behavior \cite{Ryd-bistab1}.

The dissipative case has been recently studied via a mean-field approach \cite{FullDynMF, Largecorr, cme1}, numerical calculations in one dimension \cite{tDMRG1, Ryd-QJMC1, Ryd-Numerical} and an approximate rate equation description in higher dimensions \cite{Ates06,AF-num1,Hofmann2013,Schonleber14}. These investigations highlighted the presence of various stationary regimes and the existence of first and second order phase transitions. In addition, experiments have started to probe the static and dynamic features of these systems revealing a bimodal behavior of the excitation density \cite{Experiment2} and the presence of an optical bistability \cite{Experiment1}.

The aim of this work is to shed light on the bistable transition in a dissipative Rydberg gas with particular focus on its dynamics and to connect the findings to recent experimental studies. For the stationary state, the transition is related to the spontaneous breaking of a $\Z_2$ symmetry and falls into the Ising universality class. The effective static order parameter is an appropriately shifted Rydberg excitation density. The dynamics is found to be of ``model A'' type according to the standard classification of Ref.~\cite{HH}, i.e., it is akin to a classical Ising model subject to a noise-induced spin-flipping process which does not preserve the total magnetization. However, within the dynamical framework it becomes clear that the dynamical order parameter is not formally identical to the Rydberg excitation density and the $\Z_2$ symmetry identified in the static case must be non-trivially generalized. Linking to recent experimental studies \cite{Experiment1}, we note that the dynamic transitions observed there take in fact place near the so-called \emph{spinodal lines} of the mean-field phase diagram. The connection established to ``model A'' physics allows us, moreover, to extract a universal scaling law for relaxation times for which quantitative agreement with experiment is found. We believe that this perspective will be useful for analyzing and understanding the dynamical phenomena observed in other related experiments, such as the one presented in Ref. \cite{Experiment2}.

\textit{The model.---} We employ the standard description of a Rydberg gas in terms of (fictitious) interacting spin-$1/2$ particles, where the states $\left|\downarrow\right>$ and $\left|\uparrow\right>$ correspond to the atomic ground and Rydberg states respectively.
%are identified as atomic ground state and Rydberg state, respectively.
The many-body dynamics of the system's density matrix $\hrho$ is governed by the quantum master equation (QME) $\partial_t \hrho = -i\comm{H}{\hrho \,} + \lt \mal{L}_1 +  \mal{L}_2 \rt \lqq \, \hrho \, \rqq$
%\begin{equation}
%		\partial_t \hrho = -i\comm{H}{\hrho \,} + \lt \mal{L}_1 +  \mal{L}_2 \rt \lqq \, \hrho \, \rqq,
%\label{eq:QME}
%\end{equation}
with Hamiltonian
\begin{equation}
	H = \Omega \suml{k}{} \sx_k + \Delta \suml{k}{}\wh{n}_k + \suml{k \neq p}{} V_{kp} \wh{n}_k \wh{n}_p \, .
\label{eq:H}
\end{equation}
Here $\Omega$ is the Rabi frequency and $\Delta$ the detuning of the excitation laser with respect to the ground state -- Rydberg state transition. The interaction between two atoms positioned at $\vec{r}_k$ and $\vec{r}_p$ is, e.g., of van der Waals type $V_{kp} = C_6 \abs{\vec{r}_k - \vec{r}_p}^{-6}$. Moreover, we have defined the excitation density $\wh{n}_k = (\id_k + \sz_k)/2$, with $\left\{ \sx_k, \sy_k, \sz_k   \right\}$ being the usual quantum spin operators acting on the $k$-th site. Dissipation within our model is described by the dissipator of Lindblad form
\begin{equation}
	\mal{L}_{j} \lqq (\cdot) \rqq = \suml{k}{} \lqq  L_{jk} (\cdot) L_{jk}^\dag - \frac{1}{2} \acomm{L_{jk}^\dag L_{jk}}{(\cdot) }  \rqq .
\end{equation}
Relating to previous experimental observations \cite{Rydberg2, Dephasing}, we account for two dissipation mechanisms: One is independent atomic decay (at rate $\Decay$) from the Rydberg state to the ground state, with the corresponding jump operator being $L_{1k} =\sqrt{\Decay} \, \wh{\sigma}^-_k = \sqrt{\Decay} \, (\sx_k - i \sy_k)$. The second one is dephasing of the Rydberg state relative to the ground state, occurring at rate $\Deph$ with $L_{2k} = \sqrt{\Deph} \, \wh{n}_k$.

\textit{Mean-field equations of motion.---} A mean-field treatment of the Rydberg gas has been already conducted to some extent in other works, see, e.g., \cite{FullDynMF}; here we just briefly summarize the derivation of the equations of motion. We consider the complete set of one-atom observables $\left\{ \id_k, \sx_k,\sy_k, \wh{n}_k  \right\}$ and calculate their respective averages $\left\{ 1, \vec{S} \right\} \equiv \left\{ 1, S^x, S^y, n  \right\}$ according to $\av{(\cdot)} = \trace{\hrho (\cdot)}$. Applying the QME, assuming spatial uniformity and factorising all quadratic expectations yields the closed set of dynamical equations
\begin{equation}
	\sysb{l}
		\dot{S^x} = - (\Delta + V n) S^y - \frac{\Decay + \Deph}{2} S^x  \\[3 mm]
		\dot{S^y} = 2\Omega - 4\Omega n + (\Delta + V n) S^x -  \frac{\Decay + \Deph}{2} S^y          \\[3 mm]
		\dot{n} = \Omega S^y - \Decay n   ,
	\syse  	
	\label{eq:MFE}
\end{equation}
with $V = 2 \sum_p V_{kp}$ the mean-field interaction energy.
%%%%%%%%%%%%%%%%%%%%%%%%%%%%%%%%%%%
%%%%%%%%%%%%%%%%%%%%%%%%%%%%%%%%%%%
\begin{figure}
\includegraphics[width=\columnwidth]{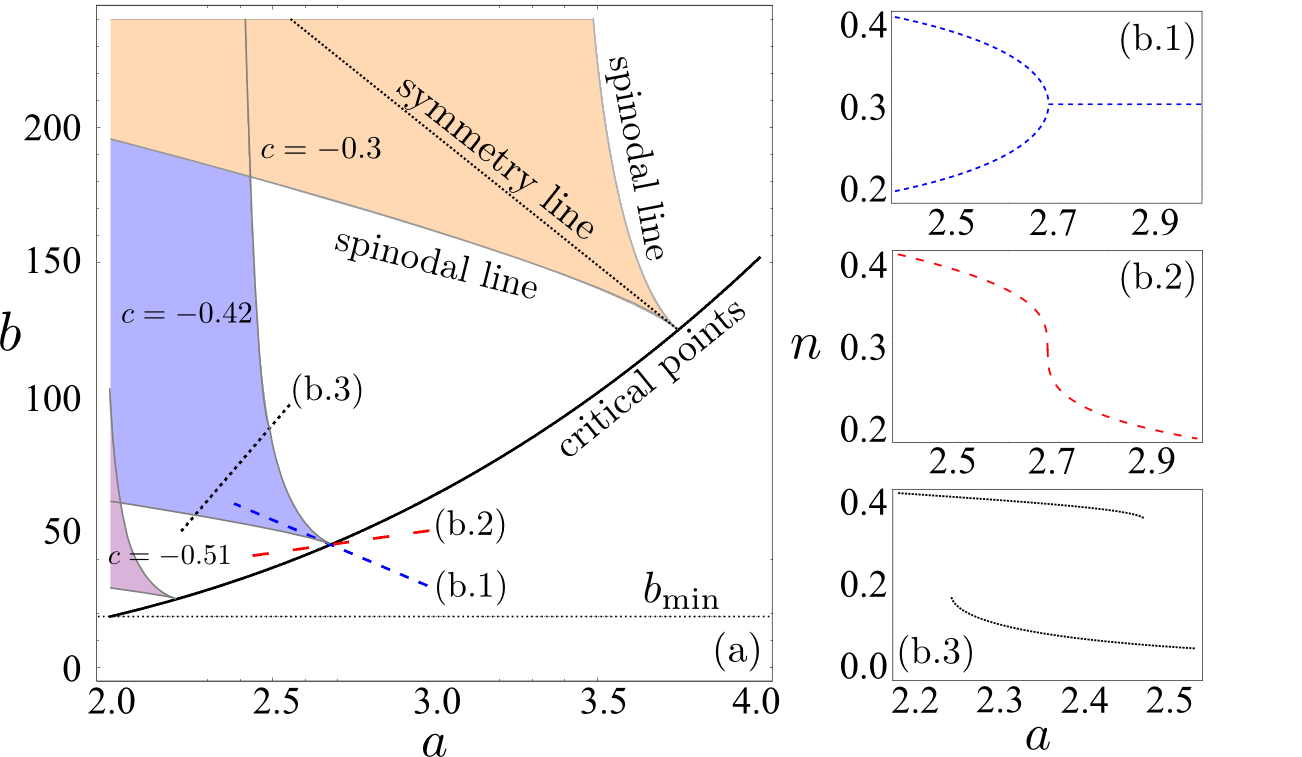}
\caption{(Color online) Phase diagram in the $a-b$ plane [as defined in Eq.~\reff{eq:effpar}] for three different values of $c$. The shaded areas correspond to domains portraying three stationary real solutions. Their boundaries identify the spinodal lines. The black curve represents the path threaded by the critical point when varying $c$, corresponding to the projection of the critical line $\left\{ a_c, b_c, c  \right\} = \left\{ -9/(8c), -27/(8c^3), c  \right\}$ onto the $a-b$ plane. The horizontal dotted line is the lower bound $b_\text{min} = 512/27$. Panels (b) relate to $c = -0.42$ and show the excitation density $n$ taken along the three cuts shown in panel (a): in panels (b.1) and (b.2) we show $n$ as observed on the blue and red dashed lines, respectively, which correspond to the ``thermal'' and ``magnetic'' directions (see main text). The black dashed line which crosses the spinodal boundaries probes instead the stable-bistable transition, corresponding to the hysteresis-like profile in panel (b.3).
}
		\label{fig:PD}
	\end{figure}
%%%%%%%%%%%%%%%%%%%%%%%%%%%%%%%%%%%
%%%%%%%%%%%%%%%%%%%%%%%%%%%%%%%%%%%

\textit{Stationary regime.---} Introducing the effective parameters
\begin{equation}
	a = 2 + \frac{1}{4} \frac{\Decay (\Decay + \Deph)}{\Omega^2} \,\, , \,\, b = \lt \frac{V}{\Omega} \rt^2 \!\!  \frac{\Decay}{\Decay + \Deph} \,\, , \,\, c= \frac{\Delta}{V}
\label{eq:effpar}
\end{equation}
allows us to formulate the problem in a concise way.  We can eliminate $S^x$ and $S^y$ from the stationary solutions of (\ref{eq:MFE}), thus obtaining an algebraic equation for the stationary average number of excitations $n$,\begin{equation}
	n \lqq a + b\lt c+n\rt^2 \rqq = 1.
	\label{eq:stateq}
\end{equation}
This expression is a cubic real polynomial in $n$ and therefore the number of real roots may vary from $1$ to $3$ depending on the specific values taken by $(a,b,c)$ within the physically allowed space $\left\{ a \geq 2, \, b\geq 0 \right\}$. In Fig.~\ref{fig:PD} we report the corresponding phase diagram in the $a-b$ plane for different choices of $c$.  The stable phase of the system corresponds to the parameter domain displaying only one acceptable solution. Complementary to this domain is the bistable regime \cite{Ryd-bistab1,FullDynMF,Ryd-int}, with Eq.~(\ref{eq:stateq}) featuring three solutions, only two of which are stable. The boundaries between stable and bimodal regimes are the spinodal lines, where at least two solutions coincide. For any value of $c$, the spinodal lines coalesce into a critical point identified by $a_c = -9/(8c)$ and $b_c = -27/(8c^3)$, which corresponds to having three coincident real solutions for Eq.~\reff{eq:stateq}. This point moves along the curve $b = (4a/3)^{3}$ shown in Fig.~\ref{fig:PD} and lies within the aforementioned physical parameter space only when $-9/16 \leq c \leq 0$. This results in a constraint on the physical parameters: a minimal value $b_\text{min} = 512/27$. From them, one can work out the threshold value $V_{\mathrm{min}} = 4(\Decay + \Deph)$ below which the transition cannot be found by just varying the laser parameters $\Omega$, $\Delta$.
%%%%%%%%%%%%%%%%%%
%%%%%%%%%%%%%%%%%%
%%%%%%%%%%%%%%%%%%
\begin{figure}[h]
	\includegraphics[width=80mm]{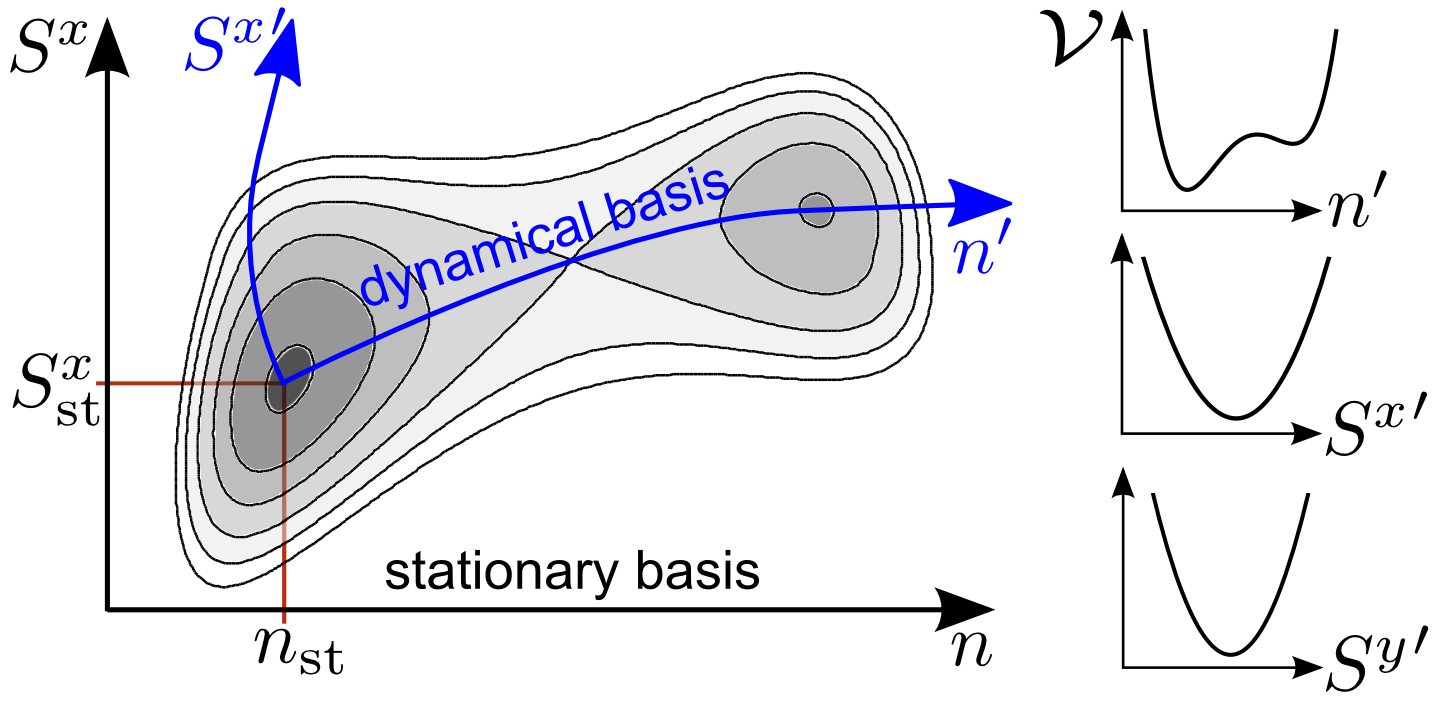}
\caption{(Color online) Stylized sketch of the mean-field potential $\mathcal{V}$, which gives rise to the dynamics governed by Eqs. (\ref{eq:MFE}), in the $S^x - n$ plane. The transformation from the stationary basis of observables $\left\{ S^x, S^y, n \right\}$ to the dynamical one ($\left\{ {S^x}', {S^y}', n' \right\}$), where the $\Z_2$ symmetry becomes manifest also in the dynamic structure, is in general nonlinear. On the right we show stylized sections of this potential obtained by moving along the directions $n'$, ${S^x}'$ and ${S^{y}}'$ respectively. Crucially, the double-well structure (responsible for the "model A" physics) is only felt by the critical $n'$, whereas along the other "massive" directions the system only probes single quadratic wells which play no role in the transition.}
\label{fig:doublewell}
\end{figure}
%%%%%%%%%%%%%%%%%%
%%%%%%%%%%%%%%%%%%

We investigate now the universal features near the critical point. To this end we expand Eq.~\reff{eq:stateq} to leading order in a perturbation of the parameters around their critical values: $a = a_c + \delta a$ and $b = b_c + \delta b$. We then study the corresponding variation of the stable solutions $n_\mathrm{st} = n_c + \delta n = -2c/3 + \delta n$ and identify a special direction $\delta b = (-9/c^2) \delta a$ [in the following referred to as \emph{symmetry line}, see Fig. \ref{fig:PD}(a)] along which the solution is invariant under the transformation $\delta n \to -\delta n$ (with a more complicated one holding for $S^x$ and $S^y$). Thus, a $\Z_2$ symmetry for the stationary value of the excitation density $n$ emerges, which is spontaneously broken in the bistable phase [see Fig.~\ref{fig:PD}(b.1)]. When approaching the critical point along the symmetry line we find $\delta n \sim (-\delta a)^{1/2}$.
%, which identifies the critical exponent $\beta = 1/2$. \changer{Note the correspondence with the behavior of the Ising magnetization $m \sim \abs{T - T_c}^\beta$ when varying the temperature at vanishing external field $h$.}
For any other direction [e.g., the red dashed line in Fig.~\ref{fig:PD}(a)] the system does not switch phases when crossing the critical point. The corresponding behavior, portrayed in Fig.~\ref{fig:PD}(b.2), is described by $\abs{\delta n} \sim \abs{ \delta a}^{1/3}$.
% \changer{again in analogy with the Ising case $m \sim h^{1/\delta}$}.
%
%The identification of the symmetry and the exponents allow us to
We can thus conclude that this transition belongs to the (static) Ising universality class with order parameter $\delta n = n-n_c$: In fact, the magnetization $m$ of an Ising model, as a function of the temperature $T$, the critical temperature $T_c$ and the magnetic field $h$ is known to obey $m(T,h=0) \sim \abs{T - T_c}^\beta$ and $m(T_c, h) \sim h^{1/\delta}$, with mean-field exponents $\beta = 1/2$ and $\delta =3$ \cite{Ma, PelVicari}. In analogy, we associate the symmetry line (b.1) to the thermal direction and any deviation from it to the presence of a magnetic field which explicitly breaks the $\Z_2$ symmetry.
Finally, a generic choice of the parameters will lead to probing the spinodal behavior shown in Fig.~\ref{fig:PD}(b.3), which has indeed been highlighted in previous theoretical and experimental studies \cite{Largecorr, Experiment1}.
%the found symmetry line can even be extended beyond linear order and reads $b = (-9/c^2) a$. It has the same meaning as the temperature line in the conventional Ising model and hence $\beta$ can be identified as thermal critical exponent. Deviations from this line correspond to the presence of a magnetic field which explicitly breaks the $\Z_2$ symmetry. This leads to the known exponent $1/3$.
\begin{figure*}[t]
%\begin{center}
\centering
	\includegraphics[trim = 0mm 5mm 0mm 5mm, clip, width=2\columnwidth]{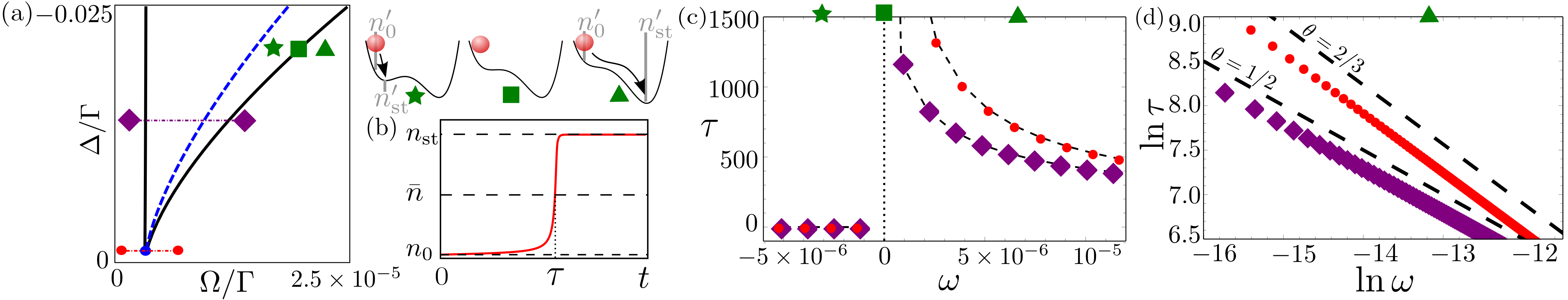}
%\end{center}
\caption{(Color online) Emergence of metastable regimes in the vicinity of the spinodal lines. In panel (a) we show the phase diagram in the $\Delta/\Decay - \Omega/\Decay$ plane. The solid curves are the spinodal lines, whereas the dashed blue one represents the symmetry line [see Fig \ref{fig:PD}(a)]. We furthermore display the qualitative structure of the mean-field potential $\mathcal{V}(n')$ for parameters corresponding to inside ($\bigstar$), on ($\blacksquare$) and outside ($\blacktriangle$) the spinodal lines (see text). In panel (b) we show an example for the relaxation of the excitation density $n$ (from an initial value $n_0$) towards the stationary value $n_\mathrm{st}$, for parameters near one of these boundaries. Here we observe a metastable plateau whose lifetime $\tau$ is determined by the first crossing time of the midpoint $\bar{n} = (n_{\text{st}} + n_0)/2$. Panel (c, d) display a power-law divergence of $\tau$ as a function of the reduced Rabi frequency $\omega = (\Omega - \Omega_c)/\Omega_c$ varied along the red and purple curves in panel (a) [see for comparison the experimental data shown in Fig. 4 of Ref. \cite{Experiment1}]. The power changes from $\theta=1/2$ in the bistable region [diamonds in panels (a,c,d)] to  $\theta=2/3$  when intersecting the critical point [disks in panels (a,c,d)].
}
\label{fig:metastab}
\end{figure*}

\textit{Dynamical vs.~static order parameter.---} The discussion so far is incomplete as it does not include the dynamical aspects of the system. As a first step, we perform an analysis of the stability of the stationary points. In their neighborhood, we expand the r.h.s. of Eqs.~(\ref{eq:MFE}) to linear order in the deviations (e.g., $\delta n = n - n_\mathrm{st}$), which obey the differential equation $ \delta \dot{\vec{S}} = M \delta \vec{S}$; the eigenvalues of the stability matrix $M$ constitute the rates of approach or escape from the stationary point.
Whenever the solution is unique, it is stable as well.  When three solutions are present, the two extremal ones are stable, while the one in the middle is unstable, cf.\ \cite{FullDynMF}.

Here, however, we focus on the universal properties that emerge near the critical point and the spinodal lines, where null eigenvalues appear. The latter are related to (leading) algebraic decays $\delta n \sim t^{-1/\zeta}$ towards stationarity. The corresponding exponent is $\zeta=1$ on the spinodal lines and $\zeta=2$ at the critical point. In the latter case, an algebraic law of the form $t^{-\beta/(\nu z)}$  is expected on the basis of scaling arguments, with $z$ being the dynamical critical exponent. The determination of the static universality class (Ising) provides us with the mean-field exponents $\beta = \nu = 1/2$. Thus, we conclude that $z=2$, describing the dynamics of a diffusive system. In addition to the null direction, the stability matrix displays two non-vanishing (``massive'') eigenvalues, which identify directions that are not involved in the critical physics (see qualitative sketch in Fig.~\ref{fig:doublewell}). Thus, the effective order parameter for the low frequency (i.e., long time) dynamics has only one component, and is described by the real variable $\delta n'$. A discrete $\Z_2$ invariance of the equations of motion under the transformation $\delta n' \to -\delta  n'$ emerges in a specific direction (i.e., the symmetry line in Fig.~\ref{fig:PD}) emanating from the critical point. This can be verified beyond the linear analysis: Applying a quadratic transformation $\delta \Sigma_i = R_{ij} \delta S_j + Q_{ijk} \delta S_j \delta S_k$, we find that along the previously found null direction, the quadratic term vanishes as well. 
%Since there is no apparent conservation law involved in the dynamics, this strongly suggests  that the dynamics of the system at hand belongs to the (one component) ``model A'' universality class. Consistently with this picture, along the symmetry line the equation of motion for the order parameter reads $\delta \dot{n}' \propto (\delta n')^3$ at leading order. In contrast, along the spinodal lines where this symmetry is not present, the leading-order equation is of the form $\delta \dot{n}' \propto (\delta n')^2$ and, consequently, governed by an exponent $\zeta=1$. 
The absence of any apparent conservation law strongly suggests that the dynamics of the system at hand belongs to the (one component) ``model A'' universality class. We remark that this is similar to the critical point in the driven open Dicke model \cite{nagy11,oztop12,DPT_Diehl,esslingerdicke3}, which however constitutes a zero dimensional model where the mean-field exponents are exact. Consistently with this picture, along the symmetry line the equation of motion reads $\delta \dot{n}' \propto (\delta n')^3$ at leading order. In contrast, along the spinodal lines where this symmetry is not present, the leading-order equation is of the form $\delta \dot{n}' \propto (\delta n')^2$ and, consequently, gives rise to an exponent $\zeta=1$. We remark that the emergent symmetry introduced above lies not among those identified in Ref.~\cite{FullDynMF} (i.e., $\left\{ \Delta, V, S^x  \right\} \to \left\{ -\Delta, -V, -S^x  \right\}$ and $\left\{ \Omega, S^x, S^y  \right\} \to \left\{ -\Omega, -S^x, -S^y  \right\}$), which are unbroken in both phases.

%Hence, the dynamical order parameter $\delta n'$ of the ``model A'' transition
%(which shows the dynamical "model A" behavior)
%does \emph{not coincide} with the static order parameter $\delta n$. Furthermore, due to the non-linearity of Eqs.~\reff{eq:MFE}, applying $R$ is not sufficient to globally identify $\delta n'$, as qualitatively sketched in Fig.~\ref{fig:doublewell}. %Although it seems unfeasible to analytically determine the latter, one can
%Still, one can proceed a step further and employ a quadratic transformation, i.e., $\delta \Sigma_i = R_{ij} \delta S_j + Q_{ijk} \delta S_j \delta S_k$. This calculation shows that along the previously found null direction, the quadratic term vanishes as well. This is in agreement with both the $\Z_2$ symmetry, which dictates that $\delta n' \to -\delta  n'$ must leave the equations invariant, and with the value of the dynamical exponent $z = 2$, which is associated to an equation that, at leading order, must read $\delta \dot{n}' \propto (\delta n')^3$. Consistently with this picture, the quadratic term remains finite along the spinodal lines, yielding a leading-order equation of the form $\delta \dot{n}' \propto (\delta n')^2$ and, consequently, an exponent $\zeta=1$.

\textit{Metastable dynamics and connection to experiments.---} We now connect these findings to recent experiments \cite{Experiment1,Experiment2} that have investigated the dynamics of dissipative Rydberg gases. The work presented in Ref.~\cite{Experiment2} has explored the phase diagram in the $\Omega-\Delta$ plane, shown in Fig.~\ref{fig:metastab}(a). Another experiment \cite{Experiment1} has demonstrated a bistable behavior similar to the one presented in Fig.~\ref{fig:PD}(b.3). Moreover, a power-law behavior of the relaxation time close to a ``critical value'' of the excitation laser strength was reported.

In order to gain some intuitive insight on the origin of these phenomena, we exploit our knowledge of the universality class and introduce a phenomenological mean-field potential $\mathcal{V}(n') = \alpha n' - \beta ( n')^2 + \gamma ( n')^4 $ which reflects the profile reported in the topmost panel on the r.h.s. of Fig.~\ref{fig:doublewell}. The corresponding mean-field dynamics is given by $ \dot{n}' = -\partial_{n'}\mathcal{V} ( n')$. For fixed $\beta > 0$ and $\gamma > 0$, this equation portrays a stable (one minimum) to bistable (two minima) transition at a threshold value $\alpha_c=(2\beta/3)^{3/2}\gamma$.
The experiments mentioned above are performed such that initially no excited atoms are present and subsequently the excitation laser is switched on at given values of $\Delta$ and $\Omega$. Within our qualitative framework, in the bistable phase ($\alpha < \alpha_c$) this may lead to a fast relaxation towards the nearest local minimum of $\mathcal{V}(n')$, which is not necessarily the global one [see $(\bigstar)$ in Fig.~\ref{fig:metastab}(a)]. Accounting for fluctuations (beyond mean-field) may in general introduce an additional time scale beyond which this picture is no longer valid and a different physics emerges. However, the agreement between our predictions and experimental observations highlighted below suggests that these features are quite robust in three dimensions and that current experiments indeed probe this ``short time physics".

When $\alpha = \alpha_c$, i.e., on a spinodal line, one of the minima becomes an inflection point [see ($\blacksquare$) in Fig.~\ref{fig:metastab}(a)]. For $\alpha \gtrsim \alpha_c$ [see ($\blacktriangle$) in Fig.~\ref{fig:metastab}(a)], in the proximity of the disappeared minimum one can identify a region of vanishing slope of the potential, which leads to a characteristic slow dynamics in a flat landscape.
This is reflected in the evolution of observables by the appearance of long-lived plateaus (see, e.g., Fig.~\ref{fig:metastab}(b) and Fig.~(4) in Ref.~\cite{Experiment1}) whose lifetime $\tau$ diverges when approaching $\alpha_c$.  By analytically solving the phenomenological equations of motion one obtains $\tau \sim \lt \alpha - \alpha_c  \rt^{-\theta}$ with $\theta = 1/2$, which agrees with the experimental estimate $\theta = 0.53 \pm 0.10$  of Ref.~\cite{Experiment1}. If, instead of a spinodal line, one crosses the critical point (i.e., $\beta = 0$ in $V(n')$), a different exponent $\theta = 2/3$ is found. It should be possible to test this prediction in the experimental setting of Ref. \cite{Experiment1}.

Note, that instead of $n'$ the standard experimental observable is the excitation density $n$. Nonetheless, critical scaling behaviors characterize the latter as well: in fact, although $n$ is a (non-linear) combination of $n'$ and the two ``massive'' variables, the latter decay exponentially fast (cfr.~Fig.~\ref{fig:doublewell}). Hence, on long time scales the dynamics is determined by the $n'$ component. This and the results of the phenomenological model are confirmed by the numerical solution of the full dynamical equations \reff{eq:MFE}: Mimicking the experimental procedure \footnote{Note that, despite the differences in the the experimental, mean-field and phenomenological approaches, which rely on varying either the laser intensity $I$, the Rabi frequency $\Omega$ or the external field $\alpha$ in $V(n')$, one can still observe the same power law. For example, $I \propto \Omega^2$, which implies $(I - I_c)^{-\theta} \propto (\Omega^2 - \Omega_c^2)^{-\theta} \sim 2\Omega_c (\Omega - \Omega_c)^{-\theta}$, with the same exponent $\theta$.}, i.e., computing $\tau$ for different values of $\Omega$ in the proximity of the spinodal lines while keeping $V, \Delta, \Decay, \Deph$ fixed [see Fig.~\ref{fig:metastab}(a)] indeed yields algebraic divergences $\tau \sim \lt \Omega - \Omega_c \rt^{-\theta}$ [see Figs.~\ref{fig:metastab}(c) and (d)] with exponents $\theta \approx 0.5$ and $\theta \approx 0.66$ for the spinodal and critical cases, respectively.

\textit{Conclusions and outlook.---} We have found strong evidence for the non-equilibrium dynamics of the dissipative Rydberg gas being governed by the "model A" universality class. This is an extensively studied class \cite{HH, Janssen, Tauber1, tauberbook, Pleimling, Baumann} and in particular it is known that the lower critical dimension is two. This would exclude the presence of a phase transition in dimension one --- a question that was raised by the authors of Ref. \cite{Largecorr} who have found numerical evidence for long-ranged correlations.

We moreover observed the emergence of a metastable regime in close proximity to the spinodal lines, whose lifetime $\tau$ diverges algebraically, in agreement with recent experimental results. Surprisingly enough, our mean-field approach is quantitatively accurate in determining the exponent of this power-law. The accord found between experimental, mean-field and phenomenological approaches suggests that this feature, even far from the critical point, is universal, i.e., it is dictated by the large-scale features of the system, rather than by its specific microscopic description. This hints at the presence of a more subtle transition taking place at the spinodal lines. The investigation of this phenomenon --- in particular in different dimensions  --- constitutes a matter of future experimental and theoretical investigation.

%\textit{Acknowledgements} -- The research leading to these results has received funding from the European Research Council under the European
%Union's Seventh Framework Programme (FP/2007-2013) /
%ERC Grant Agreement No. 335266 (ESCQUMA) and the EU-FET Grant No.
%512862 (HAIRS). We also acknowledge financial support from EPSRC Grant no.\
%EP/J009776/1.

\textit{Acknowledgements} -- The research leading to these results has received funding from the European Research Council under the European
Union's Seventh Framework Programme (FP/2007-2013) / ERC Grant Agreement No. 335266 (ESCQUMA), the EU-FET Grant No. 512862 (HAIRS), the Austrian Science Fund (FWF) through the START grant Y 581-N16  and SFB FOQUS, Project No. F4006-N16. We also acknowledge financial support from EPSRC Grant no.\
EP/J009776/1.

\bibliography{V9biblio}

\begin{thebibliography}{54}%
\makeatletter
\providecommand \@ifxundefined [1]{%
 \@ifx{#1\undefined}
}%
\providecommand \@ifnum [1]{%
 \ifnum #1\expandafter \@firstoftwo
 \else \expandafter \@secondoftwo
 \fi
}%
\providecommand \@ifx [1]{%
 \ifx #1\expandafter \@firstoftwo
 \else \expandafter \@secondoftwo
 \fi
}%
\providecommand \natexlab [1]{#1}%
\providecommand \enquote  [1]{``#1''}%
\providecommand \bibnamefont  [1]{#1}%
\providecommand \bibfnamefont [1]{#1}%
\providecommand \citenamefont [1]{#1}%
\providecommand \href@noop [0]{\@secondoftwo}%
\providecommand \href [0]{\begingroup \@sanitize@url \@href}%
\providecommand \@href[1]{\@@startlink{#1}\@@href}%
\providecommand \@@href[1]{\endgroup#1\@@endlink}%
\providecommand \@sanitize@url [0]{\catcode `\\12\catcode `\$12\catcode
  `\&12\catcode `\#12\catcode `\^12\catcode `\_12\catcode `\%12\relax}%
\providecommand \@@startlink[1]{}%
\providecommand \@@endlink[0]{}%
\providecommand \url  [0]{\begingroup\@sanitize@url \@url }%
\providecommand \@url [1]{\endgroup\@href {#1}{\urlprefix }}%
\providecommand \urlprefix  [0]{URL }%
\providecommand \Eprint [0]{\href }%
\providecommand \doibase [0]{http://dx.doi.org/}%
\providecommand \selectlanguage [0]{\@gobble}%
\providecommand \bibinfo  [0]{\@secondoftwo}%
\providecommand \bibfield  [0]{\@secondoftwo}%
\providecommand \translation [1]{[#1]}%
\providecommand \BibitemOpen [0]{}%
\providecommand \bibitemStop [0]{}%
\providecommand \bibitemNoStop [0]{.\EOS\space}%
\providecommand \EOS [0]{\spacefactor3000\relax}%
\providecommand \BibitemShut  [1]{\csname bibitem#1\endcsname}%
\let\auto@bib@innerbib\@empty
%</preamble>
\bibitem [{\citenamefont {Zinn-Justin}(2002)}]{Zinn-Justin}%
  \BibitemOpen
  \bibfield  {author} {\bibinfo {author} {\bibfnamefont {J.}~\bibnamefont
  {Zinn-Justin}},\ }\href@noop {} {\emph {\bibinfo {title} {Quantum Field
  Theory and Critical Phenomena}}},\ \bibinfo {edition} {4th}\ ed.,\
  International Series of Monographs on Physics\ (\bibinfo  {publisher}
  {Clarendon Press},\ \bibinfo {year} {2002})\BibitemShut {NoStop}%
\bibitem [{\citenamefont {Sachdev}(1999)}]{QPT}%
  \BibitemOpen
  \bibfield  {author} {\bibinfo {author} {\bibfnamefont {S.}~\bibnamefont
  {Sachdev}},\ }\href@noop {} {\emph {\bibinfo {title} {Quantum Phase
  Transitions}}}\ (\bibinfo  {publisher} {Cambridge University Press},\
  \bibinfo {year} {1999})\BibitemShut {NoStop}%
\bibitem [{\citenamefont {Ma}(1976)}]{Ma}%
  \BibitemOpen
  \bibfield  {author} {\bibinfo {author} {\bibfnamefont {S.-K.}\ \bibnamefont
  {Ma}},\ }\href@noop {} {\emph {\bibinfo {title} {Modern theory of critical
  phenomena}}},\ Frontiers in Physics\ (\bibinfo {year} {1976})\BibitemShut
  {NoStop}%
\bibitem [{\citenamefont {Hohenberg}\ and\ \citenamefont
  {Halperin}(1977)}]{HH}%
  \BibitemOpen
  \bibfield  {author} {\bibinfo {author} {\bibfnamefont {P.~C.}\ \bibnamefont
  {Hohenberg}}\ and\ \bibinfo {author} {\bibfnamefont {B.~I.}\ \bibnamefont
  {Halperin}},\ }\href@noop {} {\bibfield  {journal} {\bibinfo  {journal} {Rev.
  Mod. Phys.}\ }\textbf {\bibinfo {volume} {49}},\ \bibinfo {pages} {435}
  (\bibinfo {year} {1977})}\BibitemShut {NoStop}%
\bibitem [{\citenamefont {Kamenev}(2011)}]{kamenevbook}%
  \BibitemOpen
  \bibfield  {author} {\bibinfo {author} {\bibfnamefont {A.}~\bibnamefont
  {Kamenev}},\ }\href@noop {} {\emph {\bibinfo {title} {Field Theory of
  Non-Equilibrium Systems}}}\ (\bibinfo  {publisher} {Cambridge University
  Press},\ \bibinfo {year} {2011})\BibitemShut {NoStop}%
\bibitem [{\citenamefont {T\"auber}(2014)}]{tauberbook}%
  \BibitemOpen
  \bibfield  {author} {\bibinfo {author} {\bibfnamefont {U.~C.}\ \bibnamefont
  {T\"auber}},\ }\href@noop {} {\emph {\bibinfo {title} {Critical Dynamics -- A
  field theory approach to equilibrium and non-equilibrium scaling behavior}}}\
  (\bibinfo  {publisher} {Cambridge University Press, Cambridge},\ \bibinfo
  {year} {2014})\BibitemShut {NoStop}%
\bibitem [{\citenamefont {Baumann}\ \emph {et~al.}(2009)\citenamefont
  {Baumann}, \citenamefont {Guerlin}, \citenamefont {Brennecke},\ and\
  \citenamefont {Esslinger}}]{Baumann09}%
  \BibitemOpen
  \bibfield  {author} {\bibinfo {author} {\bibfnamefont {K.}~\bibnamefont
  {Baumann}}, \bibinfo {author} {\bibfnamefont {C.}~\bibnamefont {Guerlin}},
  \bibinfo {author} {\bibfnamefont {F.}~\bibnamefont {Brennecke}}, \ and\
  \bibinfo {author} {\bibfnamefont {T.}~\bibnamefont {Esslinger}},\ }\href@noop
  {} {\bibfield  {journal} {\bibinfo  {journal} {Nature}\ }\textbf {\bibinfo
  {volume} {464}},\ \bibinfo {pages} {1301} (\bibinfo {year}
  {2009})}\BibitemShut {NoStop}%
\bibitem [{\citenamefont {Heyl}\ \emph {et~al.}(2013)\citenamefont {Heyl},
  \citenamefont {Polkovnikov},\ and\ \citenamefont {Kehrein}}]{DPT1}%
  \BibitemOpen
  \bibfield  {author} {\bibinfo {author} {\bibfnamefont {M.}~\bibnamefont
  {Heyl}}, \bibinfo {author} {\bibfnamefont {A.}~\bibnamefont {Polkovnikov}}, \
  and\ \bibinfo {author} {\bibfnamefont {S.}~\bibnamefont {Kehrein}},\ }\href
  {\doibase 10.1103/PhysRevLett.110.135704} {\bibfield  {journal} {\bibinfo
  {journal} {Phys. Rev. Lett.}\ }\textbf {\bibinfo {volume} {110}},\ \bibinfo
  {pages} {135704} (\bibinfo {year} {2013})}\BibitemShut {NoStop}%
\bibitem [{\citenamefont {Lesanovsky}\ \emph {et~al.}(2013)\citenamefont
  {Lesanovsky}, \citenamefont {van Horssen}, \citenamefont {Gu\ifmmode
  \mbox{\c{t}}\else \c{t}\fi{}\ifmmode~\u{a}\else \u{a}\fi{}},\ and\
  \citenamefont {Garrahan}}]{DPT2}%
  \BibitemOpen
  \bibfield  {author} {\bibinfo {author} {\bibfnamefont {I.}~\bibnamefont
  {Lesanovsky}}, \bibinfo {author} {\bibfnamefont {M.}~\bibnamefont {van
  Horssen}}, \bibinfo {author} {\bibfnamefont {M.}~\bibnamefont {Gu\ifmmode
  \mbox{\c{t}}\else \c{t}\fi{}\ifmmode~\u{a}\else \u{a}\fi{}}}, \ and\ \bibinfo
  {author} {\bibfnamefont {J.~P.}\ \bibnamefont {Garrahan}},\ }\href {\doibase
  10.1103/PhysRevLett.110.150401} {\bibfield  {journal} {\bibinfo  {journal}
  {Phys. Rev. Lett.}\ }\textbf {\bibinfo {volume} {110}},\ \bibinfo {pages}
  {150401} (\bibinfo {year} {2013})}\BibitemShut {NoStop}%
\bibitem [{\citenamefont {Nagy}\ \emph {et~al.}(2011)\citenamefont {Nagy},
  \citenamefont {Szirmai},\ and\ \citenamefont {Domokos}}]{nagy11}%
  \BibitemOpen
  \bibfield  {author} {\bibinfo {author} {\bibfnamefont {D.}~\bibnamefont
  {Nagy}}, \bibinfo {author} {\bibfnamefont {G.}~\bibnamefont {Szirmai}}, \
  and\ \bibinfo {author} {\bibfnamefont {P.}~\bibnamefont {Domokos}},\ }\href
  {\doibase 10.1103/PhysRevA.84.043637} {\bibfield  {journal} {\bibinfo
  {journal} {Phys. Rev. A}\ }\textbf {\bibinfo {volume} {84}},\ \bibinfo
  {pages} {043637} (\bibinfo {year} {2011})}\BibitemShut {NoStop}%
\bibitem [{\citenamefont {Oztop}\ \emph {et~al.}(2012)\citenamefont {Oztop},
  \citenamefont {Bordyuh}, \citenamefont {Mustecaplioglu},\ and\ \citenamefont
  {Tureci}}]{oztop12}%
  \BibitemOpen
  \bibfield  {author} {\bibinfo {author} {\bibfnamefont {B.}~\bibnamefont
  {Oztop}}, \bibinfo {author} {\bibfnamefont {M.}~\bibnamefont {Bordyuh}},
  \bibinfo {author} {\bibfnamefont {O.~E.}\ \bibnamefont {Mustecaplioglu}}, \
  and\ \bibinfo {author} {\bibfnamefont {H.~E.}\ \bibnamefont {Tureci}},\
  }\href@noop {} {\bibfield  {journal} {\bibinfo  {journal} {New J. Phys.}\
  }\textbf {\bibinfo {volume} {14}},\ \bibinfo {pages} {085011} (\bibinfo
  {year} {2012})}\BibitemShut {NoStop}%
\bibitem [{\citenamefont {Torre}\ \emph {et~al.}(2013)\citenamefont {Torre},
  \citenamefont {Diehl}, \citenamefont {Lukin}, \citenamefont {Sachdev},\ and\
  \citenamefont {Strack}}]{DPT_Diehl}%
  \BibitemOpen
  \bibfield  {author} {\bibinfo {author} {\bibfnamefont {E.~G.~d.}\
  \bibnamefont {Torre}}, \bibinfo {author} {\bibfnamefont {S.}~\bibnamefont
  {Diehl}}, \bibinfo {author} {\bibfnamefont {M.~D.}\ \bibnamefont {Lukin}},
  \bibinfo {author} {\bibfnamefont {S.}~\bibnamefont {Sachdev}}, \ and\
  \bibinfo {author} {\bibfnamefont {P.}~\bibnamefont {Strack}},\ }\href
  {\doibase 10.1103/PhysRevA.87.023831} {\bibfield  {journal} {\bibinfo
  {journal} {Phys. Rev. A}\ }\textbf {\bibinfo {volume} {87}},\ \bibinfo
  {pages} {023831} (\bibinfo {year} {2013})}\BibitemShut {NoStop}%
\bibitem [{\citenamefont {Brennecke}\ \emph {et~al.}(2013)\citenamefont
  {Brennecke}, \citenamefont {Mottl}, \citenamefont {Baumann}, \citenamefont
  {Landig}, \citenamefont {Donner},\ and\ \citenamefont
  {Esslinger}}]{esslingerdicke3}%
  \BibitemOpen
  \bibfield  {author} {\bibinfo {author} {\bibfnamefont {F.}~\bibnamefont
  {Brennecke}}, \bibinfo {author} {\bibfnamefont {F.~R.}\ \bibnamefont
  {Mottl}}, \bibinfo {author} {\bibfnamefont {K.}~\bibnamefont {Baumann}},
  \bibinfo {author} {\bibfnamefont {R.}~\bibnamefont {Landig}}, \bibinfo
  {author} {\bibfnamefont {T.}~\bibnamefont {Donner}}, \ and\ \bibinfo {author}
  {\bibfnamefont {T.}~\bibnamefont {Esslinger}},\ }\href@noop {} {\bibfield
  {journal} {\bibinfo  {journal} {PNAS}\ }\textbf {\bibinfo {volume} {110}},\
  \bibinfo {pages} {11763} (\bibinfo {year} {2013})}\BibitemShut {NoStop}%
\bibitem [{\citenamefont {Kessler}\ \emph {et~al.}(2012)\citenamefont
  {Kessler}, \citenamefont {Giedke}, \citenamefont {Imamoglu}, \citenamefont
  {Yelin}, \citenamefont {Lukin},\ and\ \citenamefont {Cirac}}]{Kessler12}%
  \BibitemOpen
  \bibfield  {author} {\bibinfo {author} {\bibfnamefont {E.~M.}\ \bibnamefont
  {Kessler}}, \bibinfo {author} {\bibfnamefont {G.}~\bibnamefont {Giedke}},
  \bibinfo {author} {\bibfnamefont {A.}~\bibnamefont {Imamoglu}}, \bibinfo
  {author} {\bibfnamefont {S.~F.}\ \bibnamefont {Yelin}}, \bibinfo {author}
  {\bibfnamefont {M.~D.}\ \bibnamefont {Lukin}}, \ and\ \bibinfo {author}
  {\bibfnamefont {J.~I.}\ \bibnamefont {Cirac}},\ }\href {\doibase
  10.1103/PhysRevA.86.012116} {\bibfield  {journal} {\bibinfo  {journal} {Phys.
  Rev. A}\ }\textbf {\bibinfo {volume} {86}},\ \bibinfo {pages} {012116}
  (\bibinfo {year} {2012})}\BibitemShut {NoStop}%
\bibitem [{\citenamefont {Foss-Feig}\ \emph {et~al.}(2013)\citenamefont
  {Foss-Feig}, \citenamefont {Hazzard}, \citenamefont {Bollinger},\ and\
  \citenamefont {Rey}}]{Foss-Feig13}%
  \BibitemOpen
  \bibfield  {author} {\bibinfo {author} {\bibfnamefont {M.}~\bibnamefont
  {Foss-Feig}}, \bibinfo {author} {\bibfnamefont {K.~R.~A.}\ \bibnamefont
  {Hazzard}}, \bibinfo {author} {\bibfnamefont {J.~J.}\ \bibnamefont
  {Bollinger}}, \ and\ \bibinfo {author} {\bibfnamefont {A.~M.}\ \bibnamefont
  {Rey}},\ }\href {\doibase 10.1103/PhysRevA.87.042101} {\bibfield  {journal}
  {\bibinfo  {journal} {Phys. Rev. A}\ }\textbf {\bibinfo {volume} {87}},\
  \bibinfo {pages} {042101} (\bibinfo {year} {2013})}\BibitemShut {NoStop}%
\bibitem [{\citenamefont {Diehl}\ \emph {et~al.}(2010)\citenamefont {Diehl},
  \citenamefont {Tomadin}, \citenamefont {Micheli}, \citenamefont {Fazio},\
  and\ \citenamefont {Zoller}}]{DPT4}%
  \BibitemOpen
  \bibfield  {author} {\bibinfo {author} {\bibfnamefont {S.}~\bibnamefont
  {Diehl}}, \bibinfo {author} {\bibfnamefont {A.}~\bibnamefont {Tomadin}},
  \bibinfo {author} {\bibfnamefont {A.}~\bibnamefont {Micheli}}, \bibinfo
  {author} {\bibfnamefont {R.}~\bibnamefont {Fazio}}, \ and\ \bibinfo {author}
  {\bibfnamefont {P.}~\bibnamefont {Zoller}},\ }\href {\doibase
  10.1103/PhysRevLett.105.015702} {\bibfield  {journal} {\bibinfo  {journal}
  {Phys. Rev. Lett.}\ }\textbf {\bibinfo {volume} {105}},\ \bibinfo {pages}
  {015702} (\bibinfo {year} {2010})}\BibitemShut {NoStop}%
\bibitem [{\citenamefont {Karrasch}\ and\ \citenamefont
  {Schuricht}(2013)}]{DPT3}%
  \BibitemOpen
  \bibfield  {author} {\bibinfo {author} {\bibfnamefont {C.}~\bibnamefont
  {Karrasch}}\ and\ \bibinfo {author} {\bibfnamefont {D.}~\bibnamefont
  {Schuricht}},\ }\href {\doibase 10.1103/PhysRevB.87.195104} {\bibfield
  {journal} {\bibinfo  {journal} {Phys. Rev. B}\ }\textbf {\bibinfo {volume}
  {87}},\ \bibinfo {pages} {195104} (\bibinfo {year} {2013})}\BibitemShut
  {NoStop}%
\bibitem [{\citenamefont {Sieberer}\ \emph {et~al.}(2013)\citenamefont
  {Sieberer}, \citenamefont {Huber}, \citenamefont {Altman},\ and\
  \citenamefont {Diehl}}]{sieberer13}%
  \BibitemOpen
  \bibfield  {author} {\bibinfo {author} {\bibfnamefont {L.~M.}\ \bibnamefont
  {Sieberer}}, \bibinfo {author} {\bibfnamefont {S.~D.}\ \bibnamefont {Huber}},
  \bibinfo {author} {\bibfnamefont {E.}~\bibnamefont {Altman}}, \ and\ \bibinfo
  {author} {\bibfnamefont {S.}~\bibnamefont {Diehl}},\ }\href {\doibase
  10.1103/PhysRevLett.110.195301} {\bibfield  {journal} {\bibinfo  {journal}
  {Phys. Rev. Lett.}\ }\textbf {\bibinfo {volume} {110}},\ \bibinfo {pages}
  {195301} (\bibinfo {year} {2013})}\BibitemShut {NoStop}%
\bibitem [{\citenamefont {T\"auber}\ and\ \citenamefont
  {Diehl}(2014)}]{tauber14}%
  \BibitemOpen
  \bibfield  {author} {\bibinfo {author} {\bibfnamefont {U.~C.}\ \bibnamefont
  {T\"auber}}\ and\ \bibinfo {author} {\bibfnamefont {S.}~\bibnamefont
  {Diehl}},\ }\href {\doibase 10.1103/PhysRevX.4.021010} {\bibfield  {journal}
  {\bibinfo  {journal} {Phys. Rev. X}\ }\textbf {\bibinfo {volume} {4}},\
  \bibinfo {pages} {021010} (\bibinfo {year} {2014})}\BibitemShut {NoStop}%
\bibitem [{\citenamefont {Bloch}\ \emph {et~al.}(2008)\citenamefont {Bloch},
  \citenamefont {Dalibard},\ and\ \citenamefont {Zwerger}}]{Bloch}%
  \BibitemOpen
  \bibfield  {author} {\bibinfo {author} {\bibfnamefont {I.}~\bibnamefont
  {Bloch}}, \bibinfo {author} {\bibfnamefont {J.}~\bibnamefont {Dalibard}}, \
  and\ \bibinfo {author} {\bibfnamefont {W.}~\bibnamefont {Zwerger}},\
  }\href@noop {} {\bibfield  {journal} {\bibinfo  {journal} {Rev. Mod. Phys.}\
  }\textbf {\bibinfo {volume} {80}},\ \bibinfo {pages} {885} (\bibinfo {year}
  {2008})}\BibitemShut {NoStop}%
\bibitem [{\citenamefont {Greiner}\ \emph
  {et~al.}(2002{\natexlab{a}})\citenamefont {Greiner}, \citenamefont {Mandel},
  \citenamefont {Esslinger}, \citenamefont {H\"ansch},\ and\ \citenamefont
  {Bloch}}]{Exp-Bose-Hubbard-3D}%
  \BibitemOpen
  \bibfield  {author} {\bibinfo {author} {\bibfnamefont {M.}~\bibnamefont
  {Greiner}}, \bibinfo {author} {\bibfnamefont {O.}~\bibnamefont {Mandel}},
  \bibinfo {author} {\bibfnamefont {T.}~\bibnamefont {Esslinger}}, \bibinfo
  {author} {\bibfnamefont {T.~W.}\ \bibnamefont {H\"ansch}}, \ and\ \bibinfo
  {author} {\bibfnamefont {I.}~\bibnamefont {Bloch}},\ }\href@noop {}
  {\bibfield  {journal} {\bibinfo  {journal} {Nature}\ }\textbf {\bibinfo
  {volume} {415}},\ \bibinfo {pages} {39} (\bibinfo {year}
  {2002}{\natexlab{a}})}\BibitemShut {NoStop}%
\bibitem [{\citenamefont {Greiner}\ \emph
  {et~al.}(2002{\natexlab{b}})\citenamefont {Greiner}, \citenamefont {Mandel},
  \citenamefont {H\"ansch},\ and\ \citenamefont {Bloch}}]{Greiner}%
  \BibitemOpen
  \bibfield  {author} {\bibinfo {author} {\bibfnamefont {M.}~\bibnamefont
  {Greiner}}, \bibinfo {author} {\bibfnamefont {O.}~\bibnamefont {Mandel}},
  \bibinfo {author} {\bibfnamefont {T.~W.}\ \bibnamefont {H\"ansch}}, \ and\
  \bibinfo {author} {\bibfnamefont {I.}~\bibnamefont {Bloch}},\ }\href@noop {}
  {\bibfield  {journal} {\bibinfo  {journal} {Nature}\ }\textbf {\bibinfo
  {volume} {419}},\ \bibinfo {pages} {51} (\bibinfo {year}
  {2002}{\natexlab{b}})}\BibitemShut {NoStop}%
\bibitem [{\citenamefont {Kinoshita}\ \emph {et~al.}(2006)\citenamefont
  {Kinoshita}, \citenamefont {Wenger},\ and\ \citenamefont
  {Weiss}}]{Kinoshita}%
  \BibitemOpen
  \bibfield  {author} {\bibinfo {author} {\bibfnamefont {T.}~\bibnamefont
  {Kinoshita}}, \bibinfo {author} {\bibfnamefont {T.}~\bibnamefont {Wenger}}, \
  and\ \bibinfo {author} {\bibfnamefont {D.~S.}\ \bibnamefont {Weiss}},\
  }\href@noop {} {\bibfield  {journal} {\bibinfo  {journal} {Nature}\ }\textbf
  {\bibinfo {volume} {440}},\ \bibinfo {pages} {900} (\bibinfo {year}
  {2006})}\BibitemShut {NoStop}%
\bibitem [{\citenamefont {Cheneau}\ \emph {et~al.}(2012)\citenamefont
  {Cheneau}, \citenamefont {Barmettler}, \citenamefont {Poletti}, \citenamefont
  {Endres}, \citenamefont {Schau\ss}, \citenamefont {Fukuhara}, \citenamefont
  {Gross}, \citenamefont {Bloch}, \citenamefont {Kollath},\ and\ \citenamefont
  {Kuhr}}]{Light-cone-exp}%
  \BibitemOpen
  \bibfield  {author} {\bibinfo {author} {\bibfnamefont {M.}~\bibnamefont
  {Cheneau}}, \bibinfo {author} {\bibfnamefont {P.}~\bibnamefont {Barmettler}},
  \bibinfo {author} {\bibfnamefont {D.}~\bibnamefont {Poletti}}, \bibinfo
  {author} {\bibfnamefont {M.}~\bibnamefont {Endres}}, \bibinfo {author}
  {\bibfnamefont {P.}~\bibnamefont {Schau\ss}}, \bibinfo {author}
  {\bibfnamefont {T.}~\bibnamefont {Fukuhara}}, \bibinfo {author}
  {\bibfnamefont {C.}~\bibnamefont {Gross}}, \bibinfo {author} {\bibfnamefont
  {I.}~\bibnamefont {Bloch}}, \bibinfo {author} {\bibfnamefont
  {C.}~\bibnamefont {Kollath}}, \ and\ \bibinfo {author} {\bibfnamefont
  {S.}~\bibnamefont {Kuhr}},\ }\href@noop {} {\bibfield  {journal} {\bibinfo
  {journal} {Nature}\ }\textbf {\bibinfo {volume} {481}},\ \bibinfo {pages}
  {484} (\bibinfo {year} {2012})}\BibitemShut {NoStop}%
\bibitem [{\citenamefont {Hofferberth}\ \emph {et~al.}(2008)\citenamefont
  {Hofferberth}, \citenamefont {Lesanovsky}, \citenamefont {Schumm},
  \citenamefont {Imambekov}, \citenamefont {Gritsev}, \citenamefont {Demler},\
  and\ \citenamefont {Schmiedmayer}}]{Exp-Luttinger}%
  \BibitemOpen
  \bibfield  {author} {\bibinfo {author} {\bibfnamefont {S.}~\bibnamefont
  {Hofferberth}}, \bibinfo {author} {\bibfnamefont {I.}~\bibnamefont
  {Lesanovsky}}, \bibinfo {author} {\bibfnamefont {T.}~\bibnamefont {Schumm}},
  \bibinfo {author} {\bibfnamefont {A.}~\bibnamefont {Imambekov}}, \bibinfo
  {author} {\bibfnamefont {V.}~\bibnamefont {Gritsev}}, \bibinfo {author}
  {\bibfnamefont {E.}~\bibnamefont {Demler}}, \ and\ \bibinfo {author}
  {\bibfnamefont {J.}~\bibnamefont {Schmiedmayer}},\ }\href@noop {} {\bibfield
  {journal} {\bibinfo  {journal} {Nature Physics}\ }\textbf {\bibinfo {volume}
  {4}},\ \bibinfo {pages} {489} (\bibinfo {year} {2008})}\BibitemShut {NoStop}%
\bibitem [{\citenamefont {Schwarzkopf}\ \emph {et~al.}(2011)\citenamefont
  {Schwarzkopf}, \citenamefont {Sapiro},\ and\ \citenamefont
  {Raithel}}]{Schwarzkopf2011}%
  \BibitemOpen
  \bibfield  {author} {\bibinfo {author} {\bibfnamefont {A.}~\bibnamefont
  {Schwarzkopf}}, \bibinfo {author} {\bibfnamefont {R.~E.}\ \bibnamefont
  {Sapiro}}, \ and\ \bibinfo {author} {\bibfnamefont {G.}~\bibnamefont
  {Raithel}},\ }\href {\doibase 10.1103/PhysRevLett.107.103001} {\bibfield
  {journal} {\bibinfo  {journal} {Phys. Rev. Lett.}\ }\textbf {\bibinfo
  {volume} {107}},\ \bibinfo {pages} {103001} (\bibinfo {year}
  {2011})}\BibitemShut {NoStop}%
\bibitem [{\citenamefont {{L\"ow}}\ \emph {et~al.}(2012)\citenamefont
  {{L\"ow}}, \citenamefont {Weimer}, \citenamefont {Nipper}, \citenamefont
  {Balewski}, \citenamefont {Butscher}, \citenamefont {{B\"uchler}},\ and\
  \citenamefont {Pfau}}]{Rydberg2}%
  \BibitemOpen
  \bibfield  {author} {\bibinfo {author} {\bibfnamefont {R.}~\bibnamefont
  {{L\"ow}}}, \bibinfo {author} {\bibfnamefont {H.}~\bibnamefont {Weimer}},
  \bibinfo {author} {\bibfnamefont {J.}~\bibnamefont {Nipper}}, \bibinfo
  {author} {\bibfnamefont {J.~B.}\ \bibnamefont {Balewski}}, \bibinfo {author}
  {\bibfnamefont {B.}~\bibnamefont {Butscher}}, \bibinfo {author}
  {\bibfnamefont {H.~P.}\ \bibnamefont {{B\"uchler}}}, \ and\ \bibinfo {author}
  {\bibfnamefont {T.}~\bibnamefont {Pfau}},\ }\href@noop {} {\bibfield
  {journal} {\bibinfo  {journal} {J. Phys. B: At. Mol. Opt. Phys.}\ }\textbf
  {\bibinfo {volume} {45}},\ \bibinfo {pages} {113001} (\bibinfo {year}
  {2012})}\BibitemShut {NoStop}%
\bibitem [{\citenamefont {Saffman}\ \emph {et~al.}(2010)\citenamefont
  {Saffman}, \citenamefont {Walker},\ and\ \citenamefont {M\o{}lmer}}]{Ryd-QI}%
  \BibitemOpen
  \bibfield  {author} {\bibinfo {author} {\bibfnamefont {M.}~\bibnamefont
  {Saffman}}, \bibinfo {author} {\bibfnamefont {T.~G.}\ \bibnamefont {Walker}},
  \ and\ \bibinfo {author} {\bibfnamefont {K.}~\bibnamefont {M\o{}lmer}},\
  }\href {\doibase 10.1103/RevModPhys.82.2313} {\bibfield  {journal} {\bibinfo
  {journal} {Rev. Mod. Phys.}\ }\textbf {\bibinfo {volume} {82}},\ \bibinfo
  {pages} {2313} (\bibinfo {year} {2010})}\BibitemShut {NoStop}%
\bibitem [{\citenamefont {Malossi}\ \emph {et~al.}(2013)\citenamefont
  {Malossi}, \citenamefont {Valado}, \citenamefont {Scotto}, \citenamefont
  {Huillery}, \citenamefont {Pillet}, \citenamefont {Ciampini}, \citenamefont
  {Arimondo},\ and\ \citenamefont {Morsch}}]{Experiment2}%
  \BibitemOpen
  \bibfield  {author} {\bibinfo {author} {\bibfnamefont {N.}~\bibnamefont
  {Malossi}}, \bibinfo {author} {\bibfnamefont {M.~M.}\ \bibnamefont {Valado}},
  \bibinfo {author} {\bibfnamefont {S.}~\bibnamefont {Scotto}}, \bibinfo
  {author} {\bibfnamefont {P.}~\bibnamefont {Huillery}}, \bibinfo {author}
  {\bibfnamefont {P.}~\bibnamefont {Pillet}}, \bibinfo {author} {\bibfnamefont
  {D.}~\bibnamefont {Ciampini}}, \bibinfo {author} {\bibfnamefont
  {E.}~\bibnamefont {Arimondo}}, \ and\ \bibinfo {author} {\bibfnamefont
  {O.}~\bibnamefont {Morsch}},\ }\href@noop {} {\bibfield  {journal} {\bibinfo
  {journal} {arXiv:cond-mat/1308.1854}\ } (\bibinfo {year} {2013})},\ \Eprint
  {http://arxiv.org/abs/1308.1854} {arXiv:1308.1854 [cond-mat.quant-gas]}
  \BibitemShut {NoStop}%
\bibitem [{\citenamefont {Carr}\ \emph {et~al.}(2013)\citenamefont {Carr},
  \citenamefont {Ritter}, \citenamefont {Wade}, \citenamefont {Adams},\ and\
  \citenamefont {Weatherill}}]{Experiment1}%
  \BibitemOpen
  \bibfield  {author} {\bibinfo {author} {\bibfnamefont {C.}~\bibnamefont
  {Carr}}, \bibinfo {author} {\bibfnamefont {R.}~\bibnamefont {Ritter}},
  \bibinfo {author} {\bibfnamefont {C.~G.}\ \bibnamefont {Wade}}, \bibinfo
  {author} {\bibfnamefont {C.~S.}\ \bibnamefont {Adams}}, \ and\ \bibinfo
  {author} {\bibfnamefont {K.~J.}\ \bibnamefont {Weatherill}},\ }\href@noop {}
  {\bibfield  {journal} {\bibinfo  {journal} {Phys. Rev. Lett.}\ }\textbf
  {\bibinfo {volume} {111}},\ \bibinfo {pages} {113901} (\bibinfo {year}
  {2013})}\BibitemShut {NoStop}%
\bibitem [{\citenamefont {Hofmann}\ \emph {et~al.}(2013)\citenamefont
  {Hofmann}, \citenamefont {G\"unter}, \citenamefont {Schempp}, \citenamefont
  {Robert-de Saint-Vincent}, \citenamefont {G\"arttner}, \citenamefont {Evers},
  \citenamefont {Whitlock},\ and\ \citenamefont {Weidem\"uller}}]{Hofmann2013}%
  \BibitemOpen
  \bibfield  {author} {\bibinfo {author} {\bibfnamefont {C.~S.}\ \bibnamefont
  {Hofmann}}, \bibinfo {author} {\bibfnamefont {G.}~\bibnamefont {G\"unter}},
  \bibinfo {author} {\bibfnamefont {H.}~\bibnamefont {Schempp}}, \bibinfo
  {author} {\bibfnamefont {M.}~\bibnamefont {Robert-de Saint-Vincent}},
  \bibinfo {author} {\bibfnamefont {M.}~\bibnamefont {G\"arttner}}, \bibinfo
  {author} {\bibfnamefont {J.}~\bibnamefont {Evers}}, \bibinfo {author}
  {\bibfnamefont {S.}~\bibnamefont {Whitlock}}, \ and\ \bibinfo {author}
  {\bibfnamefont {M.}~\bibnamefont {Weidem\"uller}},\ }\href {\doibase
  10.1103/PhysRevLett.110.203601} {\bibfield  {journal} {\bibinfo  {journal}
  {Phys. Rev. Lett.}\ }\textbf {\bibinfo {volume} {110}},\ \bibinfo {pages}
  {203601} (\bibinfo {year} {2013})}\BibitemShut {NoStop}%
\bibitem [{\citenamefont {Weimer}\ \emph {et~al.}(2008)\citenamefont {Weimer},
  \citenamefont {L\"{o}w}, \citenamefont {Pfau},\ and\ \citenamefont
  {B\"{u}chler}}]{Weimer08}%
  \BibitemOpen
  \bibfield  {author} {\bibinfo {author} {\bibfnamefont {H.}~\bibnamefont
  {Weimer}}, \bibinfo {author} {\bibfnamefont {R.}~\bibnamefont {L\"{o}w}},
  \bibinfo {author} {\bibfnamefont {T.}~\bibnamefont {Pfau}}, \ and\ \bibinfo
  {author} {\bibfnamefont {H.~P.}\ \bibnamefont {B\"{u}chler}},\ }\href@noop {}
  {\bibfield  {journal} {\bibinfo  {journal} {Phys. Rev. Lett.}\ }\textbf
  {\bibinfo {volume} {101}},\ \bibinfo {pages} {250601} (\bibinfo {year}
  {2008})}\BibitemShut {NoStop}%
\bibitem [{\citenamefont {Pohl}\ \emph {et~al.}(2010)\citenamefont {Pohl},
  \citenamefont {Demler},\ and\ \citenamefont {Lukin}}]{Pohl2010}%
  \BibitemOpen
  \bibfield  {author} {\bibinfo {author} {\bibfnamefont {T.}~\bibnamefont
  {Pohl}}, \bibinfo {author} {\bibfnamefont {E.}~\bibnamefont {Demler}}, \ and\
  \bibinfo {author} {\bibfnamefont {M.~D.}\ \bibnamefont {Lukin}},\ }\href
  {\doibase 10.1103/PhysRevLett.104.043002} {\bibfield  {journal} {\bibinfo
  {journal} {Phys. Rev. Lett.}\ }\textbf {\bibinfo {volume} {104}},\ \bibinfo
  {pages} {043002} (\bibinfo {year} {2010})}\BibitemShut {NoStop}%
\bibitem [{\citenamefont {Bettelli}\ \emph {et~al.}(2013)\citenamefont
  {Bettelli}, \citenamefont {Maxwell}, \citenamefont {Fernholz}, \citenamefont
  {Adams}, \citenamefont {Lesanovsky},\ and\ \citenamefont
  {Ates}}]{Ryd-lattice1}%
  \BibitemOpen
  \bibfield  {author} {\bibinfo {author} {\bibfnamefont {S.}~\bibnamefont
  {Bettelli}}, \bibinfo {author} {\bibfnamefont {D.}~\bibnamefont {Maxwell}},
  \bibinfo {author} {\bibfnamefont {T.}~\bibnamefont {Fernholz}}, \bibinfo
  {author} {\bibfnamefont {C.~S.}\ \bibnamefont {Adams}}, \bibinfo {author}
  {\bibfnamefont {I.}~\bibnamefont {Lesanovsky}}, \ and\ \bibinfo {author}
  {\bibfnamefont {C.}~\bibnamefont {Ates}},\ }\href {\doibase
  10.1103/PhysRevA.88.043436} {\bibfield  {journal} {\bibinfo  {journal} {Phys.
  Rev. A}\ }\textbf {\bibinfo {volume} {88}},\ \bibinfo {pages} {043436}
  (\bibinfo {year} {2013})}\BibitemShut {NoStop}%
\bibitem [{\citenamefont {Schau{\ss}}\ \emph {et~al.}(2012)\citenamefont
  {Schau{\ss}}, \citenamefont {Cheneau}, \citenamefont {Endres}, \citenamefont
  {Fukuhara}, \citenamefont {Hild}, \citenamefont {Omran}, \citenamefont
  {Pohl}, \citenamefont {Gross}, \citenamefont {Kuhr},\ and\ \citenamefont
  {Bloch}}]{Ryd-lattice2}%
  \BibitemOpen
  \bibfield  {author} {\bibinfo {author} {\bibfnamefont {P.}~\bibnamefont
  {Schau{\ss}}}, \bibinfo {author} {\bibfnamefont {M.}~\bibnamefont {Cheneau}},
  \bibinfo {author} {\bibfnamefont {M.}~\bibnamefont {Endres}}, \bibinfo
  {author} {\bibfnamefont {T.}~\bibnamefont {Fukuhara}}, \bibinfo {author}
  {\bibfnamefont {S.}~\bibnamefont {Hild}}, \bibinfo {author} {\bibfnamefont
  {A.}~\bibnamefont {Omran}}, \bibinfo {author} {\bibfnamefont
  {T.}~\bibnamefont {Pohl}}, \bibinfo {author} {\bibfnamefont {C.}~\bibnamefont
  {Gross}}, \bibinfo {author} {\bibfnamefont {S.}~\bibnamefont {Kuhr}}, \ and\
  \bibinfo {author} {\bibfnamefont {I.}~\bibnamefont {Bloch}},\ }\href@noop {}
  {\bibfield  {journal} {\bibinfo  {journal} {Nature}\ }\textbf {\bibinfo
  {volume} {491}},\ \bibinfo {pages} {87} (\bibinfo {year} {2012})}\BibitemShut
  {NoStop}%
\bibitem [{\citenamefont {Ates}\ \emph {et~al.}(2012)\citenamefont {Ates},
  \citenamefont {Olmos}, \citenamefont {Garrahan},\ and\ \citenamefont
  {Lesanovsky}}]{Ryd-int}%
  \BibitemOpen
  \bibfield  {author} {\bibinfo {author} {\bibfnamefont {C.}~\bibnamefont
  {Ates}}, \bibinfo {author} {\bibfnamefont {B.}~\bibnamefont {Olmos}},
  \bibinfo {author} {\bibfnamefont {J.~P.}\ \bibnamefont {Garrahan}}, \ and\
  \bibinfo {author} {\bibfnamefont {I.}~\bibnamefont {Lesanovsky}},\ }\href
  {\doibase 10.1103/PhysRevA.85.043620} {\bibfield  {journal} {\bibinfo
  {journal} {Phys. Rev. A}\ }\textbf {\bibinfo {volume} {85}},\ \bibinfo
  {pages} {043620} (\bibinfo {year} {2012})}\BibitemShut {NoStop}%
\bibitem [{\citenamefont {Lesanovsky}\ and\ \citenamefont
  {Garrahan}(2013)}]{PRL-KinC}%
  \BibitemOpen
  \bibfield  {author} {\bibinfo {author} {\bibfnamefont {I.}~\bibnamefont
  {Lesanovsky}}\ and\ \bibinfo {author} {\bibfnamefont {J.~P.}\ \bibnamefont
  {Garrahan}},\ }\href@noop {} {\bibfield  {journal} {\bibinfo  {journal}
  {Phys. Rev. Lett.}\ }\textbf {\bibinfo {volume} {111}},\ \bibinfo {pages}
  {215305} (\bibinfo {year} {2013})}\BibitemShut {NoStop}%
\bibitem [{\citenamefont {Lee}\ \emph {et~al.}(2012)\citenamefont {Lee},
  \citenamefont {H\"affner},\ and\ \citenamefont {Cross}}]{Ryd-bistab1}%
  \BibitemOpen
  \bibfield  {author} {\bibinfo {author} {\bibfnamefont {T.~E.}\ \bibnamefont
  {Lee}}, \bibinfo {author} {\bibfnamefont {H.}~\bibnamefont {H\"affner}}, \
  and\ \bibinfo {author} {\bibfnamefont {M.~C.}\ \bibnamefont {Cross}},\ }\href
  {\doibase 10.1103/PhysRevLett.108.023602} {\bibfield  {journal} {\bibinfo
  {journal} {Phys. Rev. Lett.}\ }\textbf {\bibinfo {volume} {108}},\ \bibinfo
  {pages} {023602} (\bibinfo {year} {2012})}\BibitemShut {NoStop}%
\bibitem [{\citenamefont {Lee}\ \emph {et~al.}(2011)\citenamefont {Lee},
  \citenamefont {H\"affner},\ and\ \citenamefont {Cross}}]{FullDynMF}%
  \BibitemOpen
  \bibfield  {author} {\bibinfo {author} {\bibfnamefont {T.~E.}\ \bibnamefont
  {Lee}}, \bibinfo {author} {\bibfnamefont {H.}~\bibnamefont {H\"affner}}, \
  and\ \bibinfo {author} {\bibfnamefont {M.~C.}\ \bibnamefont {Cross}},\ }\href
  {\doibase 10.1103/PhysRevA.84.031402} {\bibfield  {journal} {\bibinfo
  {journal} {Phys. Rev. A}\ }\textbf {\bibinfo {volume} {84}},\ \bibinfo
  {pages} {031402} (\bibinfo {year} {2011})}\BibitemShut {NoStop}%
\bibitem [{\citenamefont {Hu}\ \emph {et~al.}(2013)\citenamefont {Hu},
  \citenamefont {Lee},\ and\ \citenamefont {Clark}}]{Largecorr}%
  \BibitemOpen
  \bibfield  {author} {\bibinfo {author} {\bibfnamefont {A.}~\bibnamefont
  {Hu}}, \bibinfo {author} {\bibfnamefont {T.~E.}\ \bibnamefont {Lee}}, \ and\
  \bibinfo {author} {\bibfnamefont {C.~W.}\ \bibnamefont {Clark}},\ }\href
  {\doibase 10.1103/PhysRevA.88.053627} {\bibfield  {journal} {\bibinfo
  {journal} {Phys. Rev. A}\ }\textbf {\bibinfo {volume} {88}},\ \bibinfo
  {pages} {053627} (\bibinfo {year} {2013})}\BibitemShut {NoStop}%
\bibitem [{\citenamefont {Jin}\ \emph {et~al.}(2013)\citenamefont {Jin},
  \citenamefont {Rossini}, \citenamefont {Fazio}, \citenamefont {Leib},\ and\
  \citenamefont {Hartmann}}]{cme1}%
  \BibitemOpen
  \bibfield  {author} {\bibinfo {author} {\bibfnamefont {J.}~\bibnamefont
  {Jin}}, \bibinfo {author} {\bibfnamefont {D.}~\bibnamefont {Rossini}},
  \bibinfo {author} {\bibfnamefont {R.}~\bibnamefont {Fazio}}, \bibinfo
  {author} {\bibfnamefont {M.}~\bibnamefont {Leib}}, \ and\ \bibinfo {author}
  {\bibfnamefont {M.~J.}\ \bibnamefont {Hartmann}},\ }\href {\doibase
  10.1103/PhysRevLett.110.163605} {\bibfield  {journal} {\bibinfo  {journal}
  {Phys. Rev. Lett.}\ }\textbf {\bibinfo {volume} {110}},\ \bibinfo {pages}
  {163605} (\bibinfo {year} {2013})}\BibitemShut {NoStop}%
\bibitem [{\citenamefont {H\"oning}\ \emph {et~al.}(2013)\citenamefont
  {H\"oning}, \citenamefont {Muth}, \citenamefont {Petrosyan},\ and\
  \citenamefont {Fleischhauer}}]{tDMRG1}%
  \BibitemOpen
  \bibfield  {author} {\bibinfo {author} {\bibfnamefont {M.}~\bibnamefont
  {H\"oning}}, \bibinfo {author} {\bibfnamefont {D.}~\bibnamefont {Muth}},
  \bibinfo {author} {\bibfnamefont {D.}~\bibnamefont {Petrosyan}}, \ and\
  \bibinfo {author} {\bibfnamefont {M.}~\bibnamefont {Fleischhauer}},\ }\href
  {\doibase 10.1103/PhysRevA.87.023401} {\bibfield  {journal} {\bibinfo
  {journal} {Phys. Rev. A}\ }\textbf {\bibinfo {volume} {87}},\ \bibinfo
  {pages} {023401} (\bibinfo {year} {2013})}\BibitemShut {NoStop}%
\bibitem [{\citenamefont {Lee}\ and\ \citenamefont {Cross}(2012)}]{Ryd-QJMC1}%
  \BibitemOpen
  \bibfield  {author} {\bibinfo {author} {\bibfnamefont {T.~E.}\ \bibnamefont
  {Lee}}\ and\ \bibinfo {author} {\bibfnamefont {M.~C.}\ \bibnamefont
  {Cross}},\ }\href {\doibase 10.1103/PhysRevA.85.063822} {\bibfield  {journal}
  {\bibinfo  {journal} {Phys. Rev. A}\ }\textbf {\bibinfo {volume} {85}},\
  \bibinfo {pages} {063822} (\bibinfo {year} {2012})}\BibitemShut {NoStop}%
\bibitem [{\citenamefont {Sch\"onleber}\ \emph
  {et~al.}(2014{\natexlab{a}})\citenamefont {Sch\"onleber}, \citenamefont
  {G\"arttner},\ and\ \citenamefont {Evers}}]{Ryd-Numerical}%
  \BibitemOpen
  \bibfield  {author} {\bibinfo {author} {\bibfnamefont {D.~W.}\ \bibnamefont
  {Sch\"onleber}}, \bibinfo {author} {\bibfnamefont {M.}~\bibnamefont
  {G\"arttner}}, \ and\ \bibinfo {author} {\bibfnamefont {J.}~\bibnamefont
  {Evers}},\ }\href {\doibase 10.1103/PhysRevA.89.033421} {\bibfield  {journal}
  {\bibinfo  {journal} {Phys. Rev. A}\ }\textbf {\bibinfo {volume} {89}},\
  \bibinfo {pages} {033421} (\bibinfo {year} {2014}{\natexlab{a}})}\BibitemShut
  {NoStop}%
\bibitem [{\citenamefont {Ates}\ \emph {et~al.}(2006)\citenamefont {Ates},
  \citenamefont {Pohl}, \citenamefont {Pattard},\ and\ \citenamefont
  {Rost}}]{Ates06}%
  \BibitemOpen
  \bibfield  {author} {\bibinfo {author} {\bibfnamefont {C.}~\bibnamefont
  {Ates}}, \bibinfo {author} {\bibfnamefont {T.}~\bibnamefont {Pohl}}, \bibinfo
  {author} {\bibfnamefont {T.}~\bibnamefont {Pattard}}, \ and\ \bibinfo
  {author} {\bibfnamefont {J.~M.}\ \bibnamefont {Rost}},\ }\href
  {http://stacks.iop.org/0953-4075/39/i=11/a=L02} {\bibfield  {journal}
  {\bibinfo  {journal} {J. Phys. B}\ }\textbf {\bibinfo {volume} {39}},\
  \bibinfo {pages} {L233} (\bibinfo {year} {2006})}\BibitemShut {NoStop}%
\bibitem [{\citenamefont {Hoening}\ \emph {et~al.}(2014)\citenamefont
  {Hoening}, \citenamefont {Abdussalam}, \citenamefont {Fleischhauer},\ and\
  \citenamefont {Pohl}}]{AF-num1}%
  \BibitemOpen
  \bibfield  {author} {\bibinfo {author} {\bibfnamefont {M.}~\bibnamefont
  {Hoening}}, \bibinfo {author} {\bibfnamefont {W.}~\bibnamefont {Abdussalam}},
  \bibinfo {author} {\bibfnamefont {M.}~\bibnamefont {Fleischhauer}}, \ and\
  \bibinfo {author} {\bibfnamefont {T.}~\bibnamefont {Pohl}},\ }\href@noop {}
  {\bibfield  {journal} {\bibinfo  {journal} {arXiv:cond-mat/1404.1281}\ }
  (\bibinfo {year} {2014})},\ \Eprint {http://arxiv.org/abs/1404.1281}
  {arXiv:1404.1281 [cond-mat.stat-mech]} \BibitemShut {NoStop}%
\bibitem [{\citenamefont {Sch\"onleber}\ \emph
  {et~al.}(2014{\natexlab{b}})\citenamefont {Sch\"onleber}, \citenamefont
  {G\"arttner},\ and\ \citenamefont {Evers}}]{Schonleber14}%
  \BibitemOpen
  \bibfield  {author} {\bibinfo {author} {\bibfnamefont {D.~W.}\ \bibnamefont
  {Sch\"onleber}}, \bibinfo {author} {\bibfnamefont {M.}~\bibnamefont
  {G\"arttner}}, \ and\ \bibinfo {author} {\bibfnamefont {J.}~\bibnamefont
  {Evers}},\ }\href {\doibase 10.1103/PhysRevA.89.033421} {\bibfield  {journal}
  {\bibinfo  {journal} {Phys. Rev. A}\ }\textbf {\bibinfo {volume} {89}},\
  \bibinfo {pages} {033421} (\bibinfo {year} {2014}{\natexlab{b}})}\BibitemShut
  {NoStop}%
\bibitem [{\citenamefont {Raitzsch}\ \emph {et~al.}(2009)\citenamefont
  {Raitzsch}, \citenamefont {Heidemann}, \citenamefont {Weimer}, \citenamefont
  {Butscher}, \citenamefont {Kollmann}, \citenamefont {L{\"{o}}w},
  \citenamefont {B{\"u}chler},\ and\ \citenamefont {Pfau}}]{Dephasing}%
  \BibitemOpen
  \bibfield  {author} {\bibinfo {author} {\bibfnamefont {U.}~\bibnamefont
  {Raitzsch}}, \bibinfo {author} {\bibfnamefont {R.}~\bibnamefont {Heidemann}},
  \bibinfo {author} {\bibfnamefont {H.}~\bibnamefont {Weimer}}, \bibinfo
  {author} {\bibfnamefont {B.}~\bibnamefont {Butscher}}, \bibinfo {author}
  {\bibfnamefont {P.}~\bibnamefont {Kollmann}}, \bibinfo {author}
  {\bibfnamefont {R.}~\bibnamefont {L{\"{o}}w}}, \bibinfo {author}
  {\bibfnamefont {H.~P.}\ \bibnamefont {B{\"u}chler}}, \ and\ \bibinfo {author}
  {\bibfnamefont {T.}~\bibnamefont {Pfau}},\ }\href@noop {} {\bibfield
  {journal} {\bibinfo  {journal} {New J. Phys.}\ }\textbf {\bibinfo {volume}
  {11}},\ \bibinfo {pages} {055014} (\bibinfo {year} {2009})}\BibitemShut
  {NoStop}%
\bibitem [{\citenamefont {Pelissetto}\ and\ \citenamefont
  {Vicari}(2002)}]{PelVicari}%
  \BibitemOpen
  \bibfield  {author} {\bibinfo {author} {\bibfnamefont {A.}~\bibnamefont
  {Pelissetto}}\ and\ \bibinfo {author} {\bibfnamefont {E.}~\bibnamefont
  {Vicari}},\ }\href@noop {} {\bibfield  {journal} {\bibinfo  {journal} {Phys.
  Rep.}\ }\textbf {\bibinfo {volume} {368}},\ \bibinfo {pages} {549} (\bibinfo
  {year} {2002})}\BibitemShut {NoStop}%
\bibitem [{Note1()}]{Note1}%
  \BibitemOpen
  \bibinfo {note} {Note that, despite the differences in the the experimental,
  mean-field and phenomenological approaches, which rely on varying either the
  laser intensity $I$, the Rabi frequency $\Omega $ or the external field
  $\alpha $ in $V(n')$, one can still observe the same power law. For example,
  $I \propto \Omega ^2$, which implies $(I - I_c)^{-\theta } \propto (\Omega ^2
  - \Omega _c^2)^{-\theta } \sim 2\Omega _c (\Omega - \Omega _c)^{-\theta }$,
  with the same exponent $\theta $.}\BibitemShut {Stop}%
\bibitem [{\citenamefont {Janssen}\ \emph {et~al.}(1989)\citenamefont
  {Janssen}, \citenamefont {Schaub},\ and\ \citenamefont
  {Schmittmann}}]{Janssen}%
  \BibitemOpen
  \bibfield  {author} {\bibinfo {author} {\bibfnamefont {H.~K.}\ \bibnamefont
  {Janssen}}, \bibinfo {author} {\bibfnamefont {B.}~\bibnamefont {Schaub}}, \
  and\ \bibinfo {author} {\bibfnamefont {B.}~\bibnamefont {Schmittmann}},\
  }\href@noop {} {\bibfield  {journal} {\bibinfo  {journal} {Z. Phys. B}\
  }\textbf {\bibinfo {volume} {73}},\ \bibinfo {pages} {539} (\bibinfo {year}
  {1989})}\BibitemShut {NoStop}%
\bibitem [{\citenamefont {T{\"a}uber}(2007)}]{Tauber1}%
  \BibitemOpen
  \bibfield  {author} {\bibinfo {author} {\bibfnamefont {U.~C.}\ \bibnamefont
  {T{\"a}uber}},\ }in\ \href {\doibase 10.1007/3-540-69684-9_7} {\emph
  {\bibinfo {booktitle} {Ageing and the Glass Transition}}},\ \bibinfo {series}
  {Lecture Notes in Physics}, Vol.\ \bibinfo {volume} {716},\ \bibinfo {editor}
  {edited by\ \bibinfo {editor} {\bibfnamefont {M.}~\bibnamefont {Henkel}},
  \bibinfo {editor} {\bibfnamefont {M.}~\bibnamefont {Pleimling}}, \ and\
  \bibinfo {editor} {\bibfnamefont {R.}~\bibnamefont {Sanctuary}}}\ (\bibinfo
  {publisher} {Springer Berlin Heidelberg},\ \bibinfo {year} {2007})\ pp.\
  \bibinfo {pages} {295--348}\BibitemShut {NoStop}%
\bibitem [{\citenamefont {Pleimling}(2004)}]{Pleimling}%
  \BibitemOpen
  \bibfield  {author} {\bibinfo {author} {\bibfnamefont {M.}~\bibnamefont
  {Pleimling}},\ }\href@noop {} {\bibfield  {journal} {\bibinfo  {journal}
  {Phys. Rev. B}\ }\textbf {\bibinfo {volume} {70}},\ \bibinfo {pages} {104401}
  (\bibinfo {year} {2004})}\BibitemShut {NoStop}%
\bibitem [{\citenamefont {Baumann}\ and\ \citenamefont
  {Pleimling}(2007)}]{Baumann}%
  \BibitemOpen
  \bibfield  {author} {\bibinfo {author} {\bibfnamefont {F.}~\bibnamefont
  {Baumann}}\ and\ \bibinfo {author} {\bibfnamefont {M.}~\bibnamefont
  {Pleimling}},\ }\href@noop {} {\bibfield  {journal} {\bibinfo  {journal}
  {Phys. Rev. B}\ }\textbf {\bibinfo {volume} {76}},\ \bibinfo {pages} {104422}
  (\bibinfo {year} {2007})}\BibitemShut {NoStop}%
\end{thebibliography}%

\end{document}